# North-South Distribution and Asymmetry of GOES SXR Flares during Solar Cycle 24

Anita Joshi* and Ramesh Chandra

**Abstract:** Here we present the results of the study of the north-south (N-S) distribution and asymmetry of GOES soft X-ray (SXR) flares during solar cycle 24. The period of study includes ascending, maximum and descending phases of the cycle. During the cycle double-peaked (2011, 2014) solar maximum has occurred. The cycle peak in the year 2011 is due to B-class flares excess activity in the northern hemisphere (NH) whereas C and M class flares excess activity in the southern hemisphere (SH) supported the second peak of the cycle in 2014. The data analysis shows that the SXR flares are more pronounced in 11 to 20 degree latitudes for each hemisphere. Cumulative values of SXR flare count show northern excess during the ascending phase of the cycle. However, in the descending phase of the cycle, southern excess occurred. In the cycle a significant SH dominated asymmetry exists. Near the maximum of the cycle, the asymmetry enhances pronouncedly and reverses in sign.

## 1 Introduction

The solar activity is not uniformly distributed over the solar disk and more events occur in one or another part of the disk. This phenomenon called asymmetry. In case of northern and southern hemispheres, it is N-S asymmetry. The solar N-S asymmetry is considered as one of the most interesting properties of solar activity and studied in many manifestations of solar activity such as flares, prominences, sunspots, filaments, etc. (Howard 1974; Roy 1977; Yadav et al. 1980; Swinson et al. 1986; Vizoso and Ballester 1987, 1990; Carbonell et al. 1993; Oliver and Ballester 1994; Joshi 1995; Javaraiah and Gokhale 1997; Verma 2000; Ataç and Özgüç 2001; Temmer et al. 2002; Joshi and Joshi 2004; Ballester et al. 2005; Javaraiah and Ulrich 2006; Zaatri et al. 2006; Temmer et al. 2006; Carbonell et al. 2007; Li et al. 2009a,b; Gigolashvili et al. 2011; Chowdhury et al. 2013; Chandra et al. 2013; Obridko et al. 2014; Joshi et al. 2015; Zhang et al. 2015; Javaraiah 2016; Bruevich and Yakunina 2017; Deng et al. 2017; Gurgenashvili et al. 2017; Xie et al. 2018; El-Borie et al. 2019; Li et al. 2019).

---

**Corresponding Author: Anita Joshi:** Department of Science and Technology, D. M. Ofice, NainiTal, 263002, India; Email: anitactdin@ gmail.com
**Ramesh Chandra:** Department of Physics, DSB Campus, Kumaun University, NainiTal, 263002, India; Email: rchandra.ntl@gmail.com

The existence of N-S asymmetry in various solar activity phenomena during the solar cycles is generally accepted. Various studies pointed out that the asymmetry enhances near the minimum of cycle, whereas the results obtained by Temmer et al. (2006) were contrary to those. Keeping all this in mind, the main objective of the paper is to study the N-S distribution and asymmetry of GOES SXR flares during current solar cycle 24. The paper has constructed in four sections. In Section 2, we have presented data analysis together with the N-S distribution and the N-S asymmetry. The results obtained have discussed in Section 3, and finally the conclusions of the paper have presented in Section 4.

## 2 Data and analysis

The data of GOES SXR flares from 01 January 2008 to 31 December 2016 (solar cycle 24) downloaded from NGDC's anonymous ftp server. The GOES (Geostationary Operational Environmental Satellite) instruments have provided continuous solar monitoring for several decades (Bornmann et al. 1996; Hill et al. 2005). On each GOES spacecraft, there are two X-ray Sensors that provide fluxes for the wavelength bands of 0.5–4 Å and 1–8 Å. The GOES SXR classes (i.e., B, C, M, X) represent the peak fluxes in the 1–8 Å. During the considered period the occurrence of 13207 SXR flares reported. We have used only those flares in our study for which heliographic coordinates (i.e., latitude and longitude) are given in the database. Due to this reason present study consists of only 5160 number of SXR flares for distribution and asymmetry time series analysis. Table 1 lists the SXR flares subdivided into B, C, M, and X class reported in the period under consideration. The table shows that during the current cycle, most of the flares belong to C-class (71.96%) whereas X-class (0.62%) is much less in number. Figure 1 shows evolution of SXR flare activity during solar cycle 24. In Figure 1, the overall (13207), total (5160), NH (2388) and SH (2772) flares counts have plotted.

**Table 1.** The Number and percentage of SXR flares for different importance class. Here percentage (%) expresses fraction of 5160 number of total SXR flares.

| Importance | No. of events | Percentage (%) |
|---|---|---|
| B | 966 | 18.72 |
| C | 3713 | 71.96 |
| M | 449 | 8.7 |
| X | 32 | 0.62 |

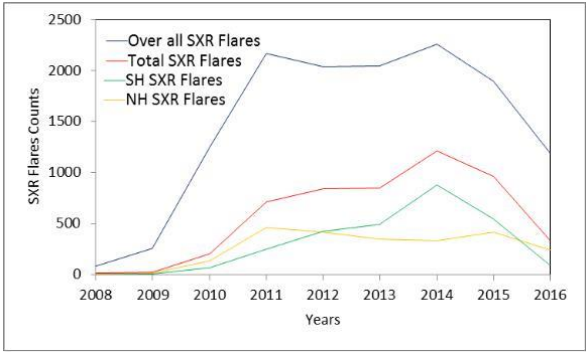

**Figure 1.** The evolution of the different considered strategies of SXR flares activity during solar cycle 24.

## 2.1 The N-S Distribution

To study the spatial distribution of SXR flares with respect to heliographic latitudes, we have evaluated the number of SXR flares in the interval of 10 degrees considering nine latitudinal belts for northern and southern hemispheres for the period 2008-2016. From the evaluation, we found that the number of flares above 40 degree latitude is very less (almost negligible) for both the hemispheres. Due to this, flares occurring above 40 degree latitude merged into one group. Therefore, the considered longitudinal belts in north and south hemispheres are 0-10, 10-20, 20-30, 30-40 and >40 degrees. In Table 2, for the years 2008-2016, the number of SXR flares at different latitudinal belts in the northern (N) and southern (S) hemispheres is listed.

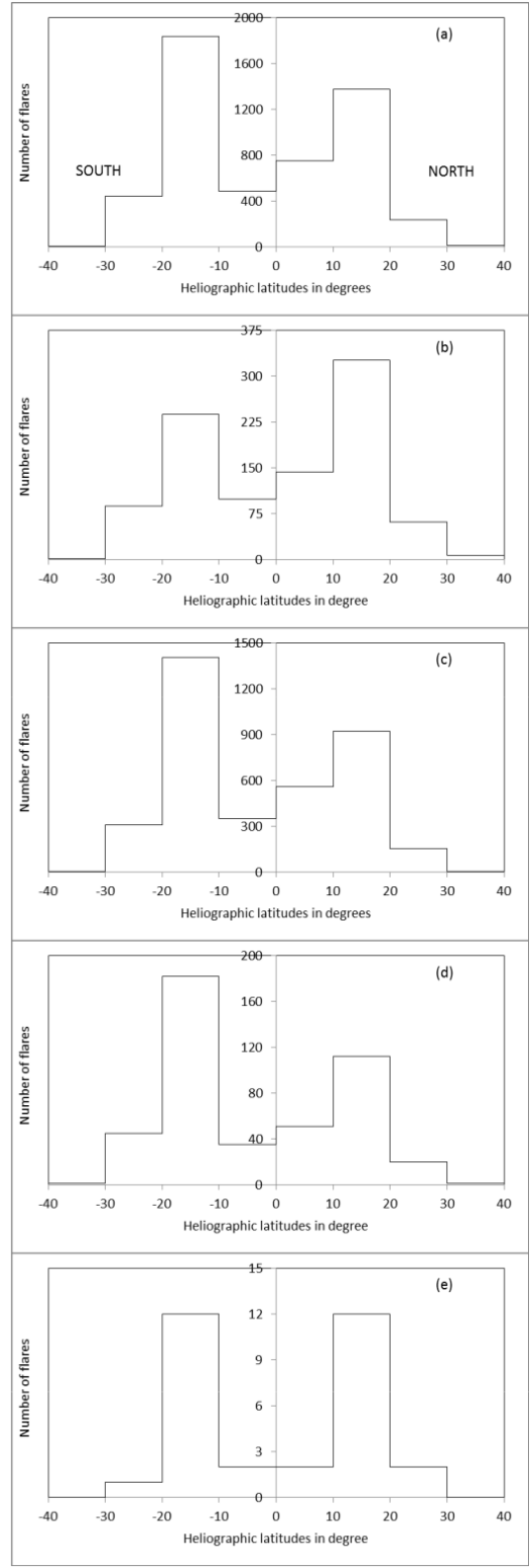

**Figure 2.** From top to bottom, plots of number of flares versus heliographic latitudes for a) Total, b) B-class, c) C-class, d) M-class and e) X-class SXR flares.

**Table 2.** Number of SXR flares in different latitudinal belts in northern (N) and southern (S) hemispheres for the years 2008-2016. The total number of flares in N and S hemispheres, the N-S asymmetry (ASY), the N-S asymmetry of the random distribution of flares (ΔASY) and the dominant hemisphere (DH) are given for each year and as well as for all latitudinal belts in the bottom of the table. Here the overall asymmetry value of −0.074 (southern dominated) was calculated using weighted average. Except in 2012 (denoted by asterisk *) all the values of asymmetry are highly significant (i.e., ASY > ΔASY).

| Years | | Number of Flares | | | | | Total | ASY | ΔASY | DH |
|---|---|---|---|---|---|---|---|---|---|---|
| | | 0–10 | 10–20 | 20–30 | 30–40 | >40 | | | | |
| 2008 | N | 1 | | | 4 | | 5 | −0.412 | −0.171 | S |
| | S | 8 | 4 | | | | 12 | | | |
| 2009 | N | | 9 | 5 | 2 | | 16 | 0.28 | 0.141 | N |
| | S | | | 8 | 1 | | 9 | | | |
| 2010 | N | 1 | 62 | 69 | 3 | 3 | 138 | 0.346 | 0.049 | N |
| | S | | 16 | 49 | 2 | | 67 | | | |
| 2011 | N | 46 | 332 | 85 | | 1 | 464 | 0.303 | 0.026 | N |
| | S | 1 | 172 | 75 | | | 248 | | | |
| 2012 | N | 110 | 268 | 36 | 4 | | 418 | −0.005* | −0.024 | S |
| | S | 8 | 304 | 108 | 2 | | 422 | | | |
| 2013 | N | 146 | 176 | 31 | | | 353 | −0.168 | −0.024 | S |
| | S | 119 | 307 | 68 | 1 | 1 | 496 | | | |
| 2014 | N | 149 | 187 | | | | 336 | −0.446 | −0.02 | S |
| | S | 183 | 613 | 81 | | | 877 | | | |
| 2015 | N | 177 | 228 | 13 | | | 418 | −0.133 | −0.023 | S |
| | S | 110 | 386 | 50 | | | 546 | | | |
| 2016 | N | 125 | 115 | | | | 240 | 0.433 | 0.039 | N |
| | S | 58 | 34 | 3 | | | 95 | | | |
| Total | N | 755 | 1377 | 239 | 13 | 4 | 2388 | −0.074 | −0.01 | S |
| | S | 487 | 1836 | 442 | 6 | 1 | 2772 | | | |
| | ASY | 0.216 | −0.143 | −0.298 | 0.368 | 0.6 | −0.074 | | | |
| | ΔASY | 0.02 | −0.012 | −0.027 | 0.162 | 0.316 | −0.01 | | | |
| | DH | N | S | S | N | N | S | | | |

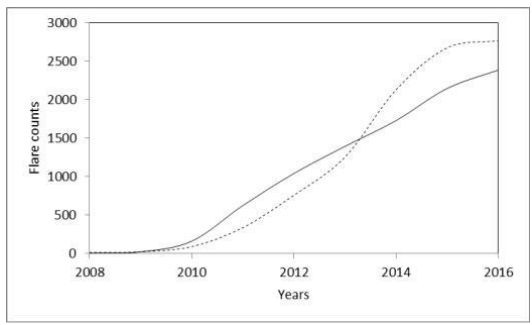

**Figure 3.** Cumulative counts of flares occurring in northern (solid line) and southern (dashed line) hemispheres.

The data analyzed in Table 2 have been plotted in Figure 2 (panel a). The figure shows total SXR flares distribution in histogram form. To show the spatial distribution changes with the intensity/flux of the flare, we have also plotted the histogram separately for the B, C, M and X class flares in panels b to e of Figure 2. To present N-S distribution in a different way, in Figure 3 the cumulative counts of northern and southern hemispheres SXR flares plotted.

## 2.2 The N-S Asymmetry

In Table 2, for the years 2008-2016, the N-S asymmetry (ASY) and the dominant hemisphere (DH) data have also given. Column 8 of Table 2 gives the yearly total number of flares in the northern (N) and southern (S) hemispheres. Using the values of column 8, N-S asymmetry index (ASY=N-S/N+S) of data values has been calculated. The calculated values of the N-S asymmetry index have been listed in Table 2 column 9 and plotted in Figure 4 (panel a). To show the statistical significance of these values we have followed Letfus (1960), Letfus and Růžičková-Topolová (1980) and Joshi (1995). Whereby the N-S asymmetry of the random distribution of SXR flares defined as ASY=±1/√[2(N+S)] and for ASY<ΔASY, ASY=ΔASY, and ASY>ΔASY the values of N-S asymmetry divided into the following three categories viz., insignificant, significant and highly significant with a low, intermediate and high probability. All the values (except in 2012) of Figure 4a are highly significant and suggesting that the N-S asymmetry is real. Table 2 and Figure 4 together indicate SH dominated asymmetry during cycle 24. Apart from this, the time span of nine years from 2008 to 2016 also shows a highly significant SH (c.f., Table 2 wherein overall asymmetry value of −0.074 is a measure of a weighted average) dominated asymmetry.

To understand how N-S asymmetry changes yearly with B, C, M and X class SXR flares the yearly asymmetry distribution of these flare events is also plotted in Figure 4 (see panels b to e). For B-class flares, Figure 4b shows a NH dominated asymmetry (all the values of the plot highly significant). For C-class flares SH dominated asymmetry occurred (except in 2008 all data values highly significant). Again, for M-class flares, Figure 4d shows a real SH dominated asymmetry (major number of data values highly significant). For X-class flares activity in both the hemispheres is equal, i.e., no hemisphere dominated (except in 2013 all data values highly significant). For B, C and M class-flares during cycle 24 an overall highly significant asymmetry corresponds to 0.117, −0.113 and −0.171. This implies that for B-class flares a real NH dominated asymmetry occurred whereas for C and M class flares SH dominated asymmetry was found which is real.

In Table 2 the ASY and DH data have also given for the latitudinal-belts. The latitudinal asymmetry data given in the bottom of Table 2 has plotted in Figure 5 (panel a). Like Table 2, the figure also shows SH dominant activity in 10-30 degree latitudinal belt and all the values of N-S asymmetry are highly significant. In panels' b to e of Figure 5, we have also plotted the latitudinal asymmetry histogram separately for the B, C, M and X class SXR flares. For B-class flares, activity is NH dominant with highly significant ( except in 20-30 degree latitudinal belt) asymmetry index. In case of C and M class flares activity is SH dominant (particularly in 10-20 and 20-30 latitudinal belts) with highly significant (except in 30-40 degree latitudinal belt) asymmetry index. For X-class flares all the values of asymmetry index are insignificant.

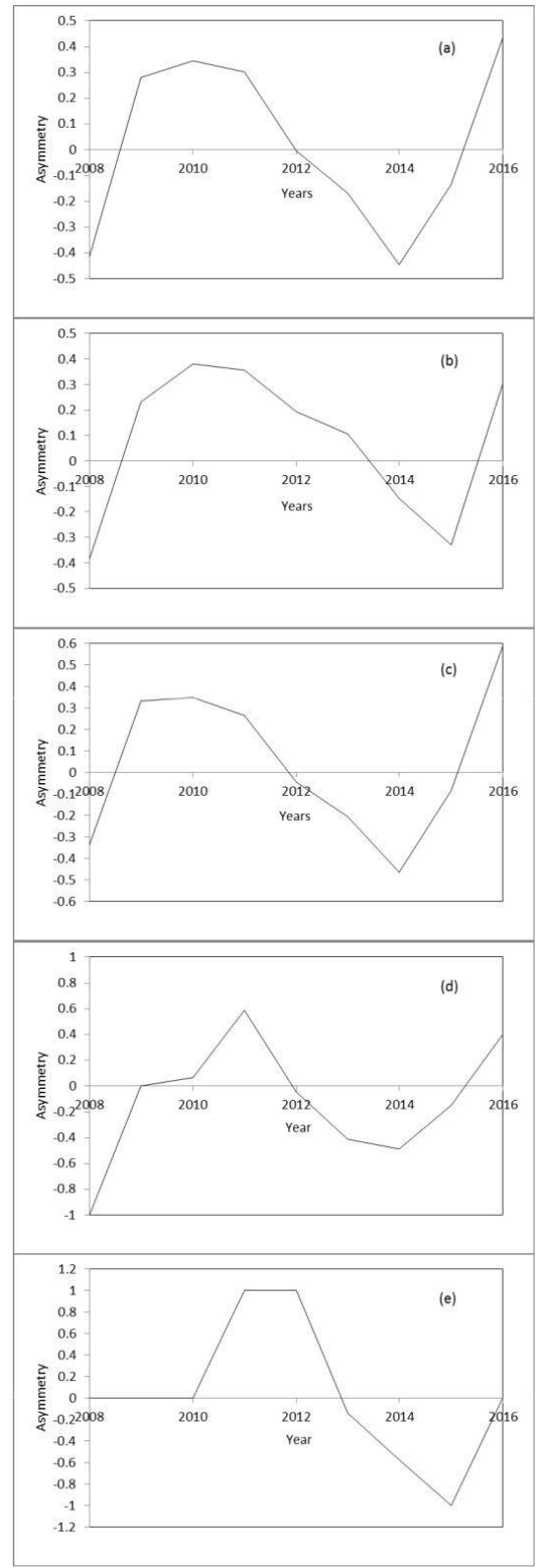

**Figure 4.** From top to bottom, plots of yearly asymmetry distribution for a) Total, b) B-class, c) C-class, d) M-class and e) X-class SXR flares.

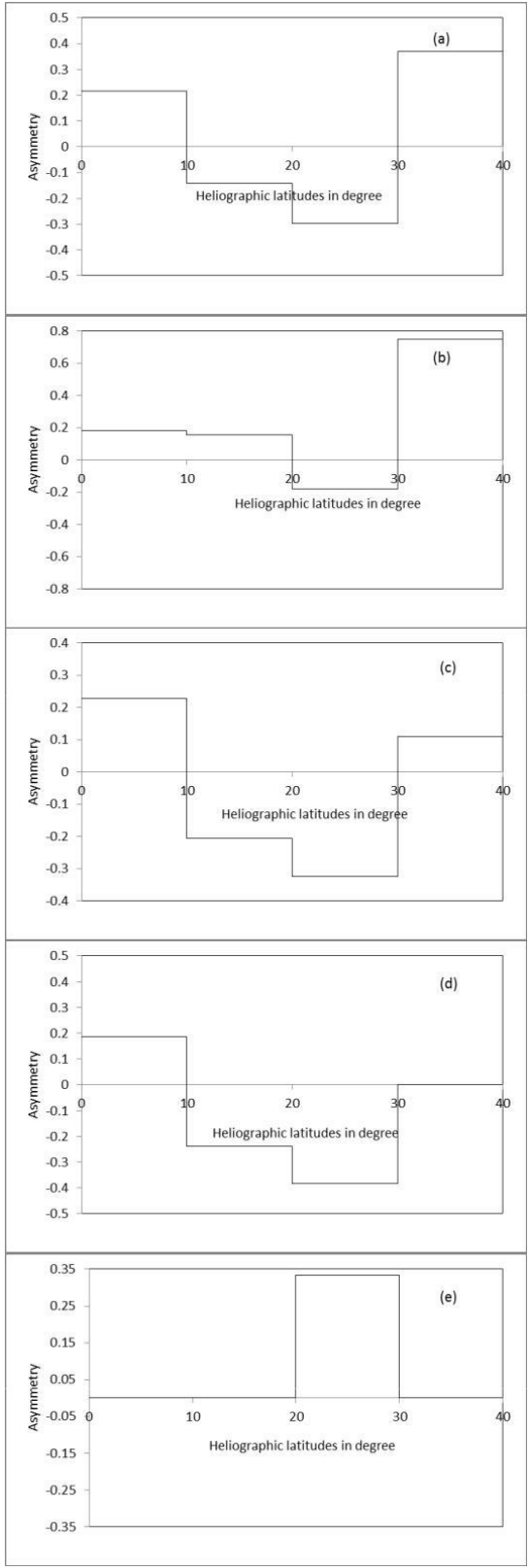

**Figure 5.** From top to bottom (a to e panels), plots of latitudinal asymmetry distribution for Total, B, C, M and X class SXR flares.

## 3 Results and discussions

We studied the N-S distribution and asymmetry of SXR flares for solar cycle 24 for the time interval of 2008 to 2016. The period of study contains the ascending-phase, the maximum-phase and the part of descending-phase, of the cycle. The cycle began in December 2008 and reached a maximum in April 2014. The cycle has a double-peaked solar maximum. The first peak occurred in 2011 and the second peak came in 2014 (see Figure 1, overall SXR flares plot). It appears from Figure 1 that the activity peak in 2011 was due to NH excess activity (number of SXR flares in NH is higher than the SH), whereas the second peak in 2014 was due to SH excess activity (number of SXR flares in SH is higher than NH). Recently, Bruevich and Yakunina (2017) have studied the N-S distribution of large ($\geq$ M1) solar flares in solar cycle 24. They have also found a strong predominance of flares in the NH in 2011 and in the SH in 2014.

The plot of overall SXR flares in Figure 1 follows the plots of solar cycle progression in other solar activity indices viz. sunspot numbers, sunspot areas, and 10.7 cm radio flux (see Hathaway 2015). The double peak structure of the cycle indicates the restructuring of the global magnetic field (Feminella and Storini 1997) and/or superposition of two oscillating processes in solar activity (Bazilevskaya et al. 2000). Recently, Kilcik and Ozguc (2014) and Javaraiah (2016) have suggested that one possible reason for a double-peaked maximum in a solar cycle is the different behavior of large and small sunspot groups, caused by a bi-dynamo mechanism (Du 2015). In the present study, the source of the double-peak behaviour is north/south asymmetry in solar activity. The activity (i.e., SXR flares) occurred in one hemisphere slightly out of phase with the activity in the other hemisphere. Due to this, early peak was associated with NH flare activity and later peak was associated with the SH flare activity. Here it seems that the source of double-peaked maximum of solar cycle 24 is different intensity class SXR flare activity. In 2011, the first peak was due to occurrence of a large number of B class flares' in NH whereas in 2014 the second peak was due to the occurrence of a large number of C and M class flares' in the SH.

Figure 2 together with Table 2 shows the maximum number of SXR flares in 11 to 20 degree latitudes for each hemisphere and as well as for all classes of SXR flares. The flares mostly occurred within ±30° latitudes. Verma et al. (1987) found that during solar cycles 19 and 20, the 11 to 20 degree latitudes are most prolific in northern and southern hemispheres for occurrence of major flares. Garcia (1990) found that during cycles 20 and 21 majorities of the flare events (M ≥ 1) occurred within ±30° latitudes. Uddin et al. (1991) studied the N-S distribution of sunspots for the cycles 20 and found that the 11 to 20 degree latitudes are most prolific in both the hemispheres. Li et al. (1998) found that during the maximum period of solar cycle 22, majorities of flares occurred in latitudes 8 to 35 degrees in both hemispheres. Verma (2000), and Joshi and Pant (2005) presented distribution of solar active prominences SAP and Hα flares and found that in both the hemispheres the SAP and flares are more prolific between 11 to 20 degree latitudes. Joshi et al. (2009, 2010) reported that for cycles 20 to 23 a larger number of SAP and SXR flares also lie in the latitudes 11 to 20 degree. Recently, Pandey et al. (2015) studied the latitudinal distribution of SXR flares and found that the 11 to 20 degree latitudes of both hemispheres are mightiest.

Figure 3 reveals that an excess of flares in the NH has developed during the rising/ascending phase of the cycle. After the maximum of the cycle, there was a rapid increase in the cumulative flare counts in the SH. Preference of the NH during the rising phase of solar cycle 24 is also reported by Chowdhury et al. (2013). For the current cycle the shift of solar activity from northern to southern hemisphere is also shown by El-Borie et al. (2019); by the 13-month moving average of sunspot numbers and sunspot areas. The cumulative behaviour of the current cycle is different from the earlier cycles (viz., 22, 23) but is quite comparable to cycle 21. Temmer et al. (2001) found that in cycle 22 there was a southern excess throughout the cycle and for cycle 23 a pronounced southern excess shifted towards NH during the initial phase of the cycle. Joshi et al. (2007) reported that during solar cycle 23 the cumulative flare counts show a southern excess in the rising phase of the cycle, while after this phase a constant northern excess prevails. Recently Joshi et al. (2015) reported that the cumulative plots of the flare index indicated slight excess of activity in NH during cycle 21 while during the cycle 22 and 23 the excess of activity occurred in SH. Temmer et al. (2001) found that for solar cycle 21 an excess of flares occurring in the NH during the rising phase of the cycle, staying roughly constant during the major activity and at the last the northern excess shifted towards southern. Joshi et al. (2015) found that for cycle 21, the cumulative values of sunspot area show a northern excess during most of the cycle, however, at the end of the cycle southern excess of sunspot area occurred.

Figure 4 together with Table 2 indicate existences of a real and highly significant SH dominant N-S asymmetry during the current solar cycle 24. The figure also shows the impact role of C-class flares in yearly asymmetry distribution of total SXR flares (cf., Table 1 also). The study has supported southern activity dominated predictions of Verma (1992) and Li et al. (2009a) for the current cycle. Zhang et al. (2015) studied N-S asymmetry of X-ray flares of class B and higher, during solar cycle 24. They found that the asymmetry was fairly constant and negative (i.e., SH dominant) and SH rotates faster than the NH. For the number of sunspot groups' data, Javaraiah (2016) also found similar trend (i.e., southern dominance activity) of N-S asymmetry in solar cycle 24. Recently, Li et al. (2019) have found that during cycle 24 for sunspot areas' dominant hemisphere was SH which suggests 12 solar cycles periodic behavior. Figure 4 shows that the asymmetry is maximum or more pronounced at the time of maximum (2014) of the cycle. This result is contradicted to various earlier studies (Swinson et al. 1986; Vizoso and Ballester 1987, 1990; Carbonell et al. 1993; Ataç and Özgüç 1996; Joshi and Joshi 2004; Li et al. 2009b). The earlier studies show that the N-S asymmetry is higher when approaching the minimum of a cycle. On the other hand, this result supports the studies of N-S asymmetry of hemispheric sunspot numbers during solar cycles 18-23 (Temmer et al. 2006). Temmer et al. (2006) found that for all the cycles the asymmetry enhanced near the cycle maximum. Figure 4 also shows reversal in the sign (positive to negative) of asymmetry during the maximum (2011-2012) of cycle which coincides very well with the magnetic field reversal of the Sun in mid-2012 (Mordvinov et al. 2016). The figure also shows a tendency to recover the positive value of the asymmetry during the decay of the cycle (2015-2016). The results strongly support findings of Vizoso and Ballester (1987).

Figure 5 presents the latitudinal distribution of asymmetry during solar cycle 24 and shows (panel a) SH dominant activity in 10-30 degree latitudinal belt (cf., Table 2) whereas in other latitudinal belts the activity is north dominated which is very less in comparison to the south. The figure also shows that C and M class flares strongly supported this behaviour of current cycle which implies that majority of flare events of C and M class occurred in 10-30 degree latitudinal belt in the south. Consequently, the C and M class flares support the southern dominated asymmetry and the second maximum of the cycle.

## 4  Conclusions

In the present paper the N-S distribution and asymmetry, of GOES SXR flares during solar cycle 24 have been studied. The cycle has a double-peaked (2011, 2014) solar maximum. The cycle peak in 2011 is due to B class flares' excess activity in the NH; whereas in the year 2014, the second peak was due to excess activity of C and M class flares in the SH. This result supported Du (2015) bi-dynamo mechanism, which indicates that the different behaviour of B, C, and M class SXR flares, caused by a bi-dynamo mechanism.

From the present study, it is found that the majority of the flare events occurred within ±30° latitudes. The latitudinal distribution of all class SXR flare events show almost identical behaviour. The flares are most prolific between 11 to 20 degree latitudes in northern and southern hemispheres. This result supported various earlier studies. The cumulative plots of flare activity indicated the excess of activity in the NH during the ascending phase of the cycle, while during the later phases of the cycle, the excess of activity shifted towards the southern hemisphere. This behaviour of the current cycle is identical to cycle 21.

In the current study, we found highly significant southern hemisphere dominated asymmetry. From the present study, we also found that in case of solar cycle 24 the N-S asymmetry is more pronounced at the time of the maximum of the cycle. The result strongly supports Temmer et al. (2006) study which is related to various earlier solar cycles. In case of the current cycle, the sign of asymmetry changes during the maximum of the cycle at the time of reversal of the Sun's magnetic dipole, which recovers the positive value quickly. The latitudinal distribution of asymmetry shows that the C and M class flare activity is dominant in 10-30 degree latitudinal belt, which also supports southern dominant asymmetry as well as the second maximum of the cycle.

**Acknowledgments:** Authors gratefully thank the anonymous referees for constructive comments and suggestions, which improved the quality and presentation of the paper. The authors also acknowledge support of National Geophysical Data Center, Boulder, Colorado, USA for providing the data of GOES SXR flares via ftp server. RC acknowledges the support from SERB- DST project no. SERB/F/7455/ 2017-17.